\documentclass[aps,prd,twocolumn,showpacs,superscriptaddress,groupedaddress,nofootinbib]{revtex4}

\usepackage[latin1]{inputenc}
\usepackage{float}
\usepackage{latexsym}
\usepackage{graphicx}
\usepackage{amssymb}
\usepackage{amsmath}
\usepackage{amsfonts}
\usepackage{hyperref}
\usepackage{cool}
\usepackage{dcolumn}
\usepackage{textcomp}
\usepackage{color}

\begin{document}

\title{Evidence for interacting dark energy from BOSS}

\author{Elisa G. M. Ferreira}\email{elisa.ferreira@mail.mcgill.ca}
\author{Jerome Quintin\footnote{Vanier Canada Graduate Scholar.}}\email{jquintin@physics.mcgill.ca}
\affiliation{Department of Physics, McGill University, Montr\'eal, Qu\'ebec, H3A 2T8, Canada}
\author{Andr\'e A. Costa}\email{alencar@if.usp.br}
\author{E. Abdalla}\email{eabdalla@usp.br}
\affiliation{Instituto de F\'isica, Universidade de S\~ao Paulo, C.P. 66318, 05315-970, S\~ao Paulo, SP, Brazil}
\author{Bin Wang}\email{wang\_b@sjtu.edu.cn}
\affiliation{Department of Physics and Astronomy, Shanghai Jiao Tong University, 200240 Shanghai, China }
\date{\today}

\begin{abstract}
The result presented
by the BOSS-SDSS Collaboration measuring the
baryon acoustic oscillations of the Lyman-alpha
forest from high-redshift quasars indicates a
$2.5\sigma$ departure from the standard
$\Lambda$-cold-dark-matter model. This is the first time that
the evolution of dark energy at high redshifts
has been measured, and the current results cannot
be explained by simple generalizations of the
cosmological constant. We show here that a simple
phenomenological interaction in the dark sector
provides a good explanation for this deviation,
naturally accommodating the Hubble parameter
obtained by BOSS, $H(z=2.34)=222 \pm 7
~\mathrm{km~s^{-1}~Mpc^{-1}}$.
By performing a global fit of the parameters with the
inclusion of this new data set together with the
Planck data for the interacting model,
we are able to show that some interacting models have
constraints for $H(2.34)$ and $D_\mathrm{A}(2.34)$ that are
compatible with the ones obtained by the BOSS
Collaboration, showing a better concordance than
$\Lambda$CDM. We also show that the interacting models that have a small positive coupling constant,
which helps alleviate the coincidence problem, are compatible with the cosmological observations.
Adding the likelihood of these new baryon
acoustic oscillations data shows an
improvement in the global fit, although it is not statistically
significant. The coupling constant could not be fully
constrained by the data sets used, but the dark energy equation
of state shows a slight preference for a value different
from a cosmological constant.
\end{abstract}


\maketitle

\section{Introduction}

One of the biggest challenges
in cosmology and astrophysics nowadays is to understand the nature
of the two most abundant components of the Universe: dark energy
and dark matter. These are usually described as two independent
components where dark matter is responsible for most of the
nonrelativistic matter in the Universe and where dark energy
is responsible for the late time acceleration of our Universe,
which is described by a cosmological constant in the
$\Lambda$-cold-dark-matter ($\Lambda$CDM)
model. This standard model is widely used to describe the cosmological
evolution of the Universe\ \cite{Ade:2013zuv},
and it fits very well the current observational data.
However, this model has some theoretical and observational challenges
(see, e.g., Ref.\ \cite{Bull:2015stt})
that open the way for alternative models of dark energy.

Recently, the Baryon Oscillation Spectroscopic Survey (BOSS)
experiment of the Sloan Digital Sky Survey (SDSS) Collaboration presented new evidence against the
$\Lambda$CDM model\ \cite{Delubac:2014aqe} based on the measurements of the
baryon acoustic oscillations (BAO) flux-correlation
function of the Lyman-alpha (Ly-$\alpha$) forest from $158,401$ quasars at high redshifts
($2.1 \leq z \leq 3.5$). Comparatively to previous experiments, they provide the line of
sight and tangential BAO components, and this allows one to determine the angular distance
and the Hubble distance independently. Their results indicate a deviation from $\Lambda$CDM
of the Hubble parameter and of angular distance at an average redshift of $2.34$
(roughly $2.5\sigma$ and $2.2\sigma$ deviations from Planck+Wilkinson Microwave Anisotropy Probe (WMAP)
polarization data and WMAP9+ACT+SPT,
respectively). Assuming a $\Lambda$CDM Universe, this implies a negative energy density
for the dark energy component, $\frac{\rho_\mathrm{DE}(z=2.34)}{\rho_\mathrm{DE}(0)}=-1.2\pm0.8$, which
is $2.5 \sigma$ away from the expected value. We point out that BOSS is not optimized to
observe quasars at such high redshifts. However, if more data or other experiments show
that this discrepancy stands, then it would indicate that $\Lambda$CDM needs to be
revised. Its simplest generalization would consist in allowing for dynamical dark energy
(see Ref.\ \cite{Copeland:2006wr} for a review), but this would not be enough to fix this
discrepancy. In dynamical dark energy models, all matter contents are individually conserved,
and so, agreeing with the BOSS result for $H(z=2.34)$ would require a negative energy density for
dark energy\ \cite{Delubac:2014aqe}.
This may lead one to study very exotic forms of dark energy.

A simpler solution is to consider interacting dark energy.
Indeed, dark energy could couple to gravity, neutrinos, or dark matter since its effects
have only been detected gravitationally. Interaction with baryonic matter (or radiation)
has very tight constraints from observations \cite{Damour:1990tw} and must be very small or negligible. In this
sense, we are interested in models in which dark energy interacts with the dark matter
component. In a field theory description of those components, this interaction is allowed
and even mandatory\ \cite{Micheletti:2009pk,Micheletti:2009jy}. However, the main
motivation to introduce such an interaction is to alleviate the coincidence problem,
which can be done given an appropriate interaction.

Since the nature of the dark sector is unknown, the study of these coupled dark energy
models is challenging. Many different models of this interaction have been studied in the
literature from the point of view of either interacting field theory or phenomenology (for
a classification of those models, see Ref.\ \cite{Koyama:2009gd}). As an example of
phenomenological study, one can consider holographic dark energy or a quintessence field
interacting with a dark matter fluid\ \cite{Amendola:1999er,Wang:2005jx,He:2008si,He:2008tn,He:2010im}. There
are also attempts to develop Lagrangian models where one postulates an interaction between
the scalar field, playing the role of dark energy, and a fermionic field, playing the role
of dark matter\ \cite{Pavan:2011xn,Micheletti:2009pk,Abdalla:2012ug,Costa:2014pba}
(see, however, Ref.\ \cite{Faraoni:2014vra}).

Recently, there have been studies of interacting dark energy models in light
of new probes\ \cite{Salvatelli:2014zta,Nunes:2016dlj,Costa:2016tpb,Marcondes:2016reb}.
However, we note that there has been only little exploration of the consequences of the
results from BOSS in the literature\ \cite{Aubourg:2014yra,Cardenas:2014jya,Sahni:2014ooa},
and these studies do not explore the idea of interacting dark energy and dark matter.
Thus, it would be interesting to see what the phenomenological implications from BOSS for
interacting dark energy are. Since this model allows for one of the components to decay into
the other, we claim that energy flow from dark energy to dark matter implies a smaller
amount of dark matter in the past, thus accommodating for the value of the Hubble
parameter at $z=2.34$ found by BOSS and still maintaining the cosmology today close to
$\Lambda$CDM. For a first test, we perform a comparison by showing that the observational value of
the Hubble parameter from quasars given by the BOSS Collaboration,
$H(2.34)=222 \pm 7 ~\mathrm{km~s^{-1}~Mpc^{-1}}$, is consistent with the interacting model
with a small positive coupling constant.
This comparison serves to indicate that the interaction is able to accommodate the BOSS Collaboration result.
After that, we perform a full Markov chain Monte Carlo (MCMC) analysis using 
the new BOSS data together with the Planck data for the interacting model.
We show that the constraints on $H(z=2.34)$ and $D_\mathrm{A}(z=2.34)$ for the interacting model
are compatible with the values obtained by the BOSS team, showing a slightly better concordance when compared to $\Lambda$CDM.

\section{Model}

\subsection{Theoretical setup}
Given the energy conservation of the full
energy-momentum tensor, we can suppose that
the fluid equations representing dark energy (DE)
and dark matter (DM) are not conserved separately.
In a Friedmann-Robertson-Walker Universe, we take
\begin{align}
\label{eq:intera_fen}
\dot{\rho}_\mathrm{DM}+3H\rho_\mathrm{DM}&=Q_\mathrm{DM}=+Q\,,\nonumber \\
\dot{\rho}_\mathrm{DE}+3H\left(1+\omega_\mathrm{DE}\right)\rho_\mathrm{DE}&=Q_\mathrm{DE}=-Q\,,
\end{align}
and all other components follow the standard conservation equations.
In the above equations, $\rho_\mathrm{DM}$ and $\rho_\mathrm{DE}$ are the energy densities
for dark matter and dark energy, respectively;
$\omega_\mathrm{DE}=p_\mathrm{DE}/\rho_\mathrm{DE}$ is the equation of
state (EoS) of dark energy, considered constant in this work; and $Q$ indicates the
interaction between dark energy and dark matter.
One can take the Taylor expansion of the general interaction term $Q(\rho_\mathrm{DM},\rho_\mathrm{DE})$,
and thus, it can be represented phenomenologically as
$Q \simeq 3H(\xi_{1}\rho_\mathrm{DM}+\xi_{2}\rho_\mathrm{DE})$, where the
coefficients $\xi _{1}$ and $\xi _{2}$ are to be
determined by observations
\cite{Feng:2008fx,He:2010im}. Following our
definition, if $Q>0$, then dark energy decays
into dark matter, and for $Q<0$, the energy flow is in
the opposite direction. The first case is consistent with the requirement that the energy
density for dark energy must be of the same order as the one for dark matter for a longer period
of time in order to alleviate the coincidence problem.

The validity of the phenomenological interacting dark energy model
was studied in Ref.\ \cite{He:2008si}, where it was found
that the curvature perturbations can always be
stable when the interaction is proportional to
the energy density of dark energy, i.e.\ when $\xi_{1}=0$
while $\xi_2\neq 0$, except when $\omega = -1$, which represents
a central singularity in the cosmological perturbation equations.
This is true for a constant EoS within the ranges $-1 <  \omega_\mathrm{DE} < 0$
(we call this model I) and $\omega_\mathrm{DE}<-1$ (we call this model II).
If the interaction term is proportional to the dark matter energy density,
i.e.~$\xi_{1}\neq 0$ while $\xi_2= 0$, then the curvature
perturbations are only stable when $\omega_\mathrm{DE}<-1$ (we call this model III).
The models are summarized in Table\ \ref{table:models}.

\begin{table}[htb]
\centering \caption{Interacting dark energy
models considered in this paper.}
\begin{tabular}{ccc}
 \toprule
    Model & $Q$ & DE EoS  \\
    \hline
    I & $3\xi _{2} H\rho_\mathrm{DE}$ & $-1 <  \omega < 0$  \\
    II & $3\xi_{2}H\rho_\mathrm{DE}$  & $\omega < -1$ \\
    III & $3\xi_{1}H\rho_\mathrm{DM}$  & $\omega < -1$ \\
    \lasthline
\end{tabular}
\label{table:models}
\end{table}

In this framework, the Friedmann equations can be written as
\begin{align} 
\label{H2} 
&H^{2}(z)=\frac{8\pi G}{3}\left[ \rho_\mathrm{DE}(z)+\rho_\mathrm{DM}(z)+\rho_{\mathrm{b}}(z) \right]\,, \\
\label{hdot}
&\dot H=-4\pi G\left[\rho_\mathrm{DM}(z)+ \rho_\mathrm{b}(z) +(1+\omega_\mathrm{DE})\rho_\mathrm{DE}(z)  \right]\,,
\end{align}
where we are considering a Universe composed of only dark energy, dark matter, and baryons ($\rho_\mathrm{b}$).
We will use these equations to construct the Hubble parameter for each of the interacting models
and compare it with the Hubble parameter inferred from the BOSS quasar data in the next subsection.

For models I and II, the energy densities for dark energy and dark matter
behave as\ \cite{He:2008tn}
\begin{align}
\label{int_DE}
\rho_\mathrm{DE}=&~(1+z)^{3\left(1+\omega_\mathrm{DE}+\xi_{2}\right)}\rho_\mathrm{DE}^0\,, \nonumber \\
\rho_\mathrm{DM}=&~(1+z)^3 \nonumber \\
  &\times\left\{\frac{\xi_{2}\left[1-(1+z)^{3(\xi_{2} +\omega_\mathrm{DE})}\right]\rho_\mathrm{DE}^0}{\xi_{2}+\omega_\mathrm{DE}}+\rho_\mathrm{DM}^0\right\}\,,
\end{align}
where the superscript 0 indicates quantities measured today.
The baryonic density is given by the standard expression,
proportional to $(1+z)^3$. For model III, the
evolution of the energy densities is given by\ \cite{He:2008tn}
\begin{align}
\label{int_DM}
\rho_\mathrm{DE}=&~(1+z)^{3(1+\omega_\mathrm{DE})} \left( \rho_\mathrm{DE}^{0} + \frac{\xi_{1} \rho_\mathrm{DM}^{0}}{\xi_{1}+\omega_\mathrm{DE}} \right) \nonumber \\
&-\frac{\xi_{1}}{\xi_{1}+\omega_\mathrm{DE}}(1+z)^{3(1-\xi_{1})} \rho_\mathrm{DM}^{0} \,, \nonumber \\
\rho_\mathrm{DM}=&~\rho_\mathrm{DM}^0 (1+z)^{3-3 \xi_{1}}\,.
\end{align}

One can see from these equations that if there is an energy flow from
dark energy to dark matter (i.e., if the coupling constant is positive),
then the energy density for dark matter is
always smaller than what one would expect in the standard $\Lambda$CDM model.
Since $\rho_\mathrm{DM}$ is the dominant contribution in the Friedmann equations
at higher redshifts and since observations indicate that the Universe is well explained
by the $\Lambda$CDM model at low redshifts (e.g., Ref.\ \cite{Ade:2013zuv}), one can see from Eq.\ \eqref{hdot} that the interaction
implies a smaller Hubble parameter in the past in comparison with $\Lambda$CDM,
when $H_0$ is held fixed and for a positive coupling constant.

Furthermore, this mildly helps alleviate the coincidence problem (the fact
that we do not understand why the energy densities of dark energy and
dark matter are so close today). As it can be seen in Ref.\ \cite{Ferreira:2013eja},
a positive coupling constant implies that the quantity
$r\equiv\rho_\mathrm{DM}/\rho_\mathrm{DE}$ decreases at a slower rate in the interacting model than in
the $\Lambda$CDM model. This makes the energy density of dark energy closer
to that of dark matter in the past,
giving us a better understanding of their closer values today.

\subsection{Hubble parameter at $z=2.34$}

In order to gain some intuition before performing the proper
statistical analysis,
let us see whether the measured
value of the Hubble parameter by the BOSS Collaboration,
$H(2.34)=222 \pm 7 ~\mathrm{km~s^{-1}~Mpc^{-1}}$, can be accommodated
by the phenomenological interacting models introduced
above. From this perspective, we compare the
Hubble parameter constructed theoretically with its observational
value at $z = 2.34$.

\begin{figure*}[htb]
\includegraphics[scale=0.78]{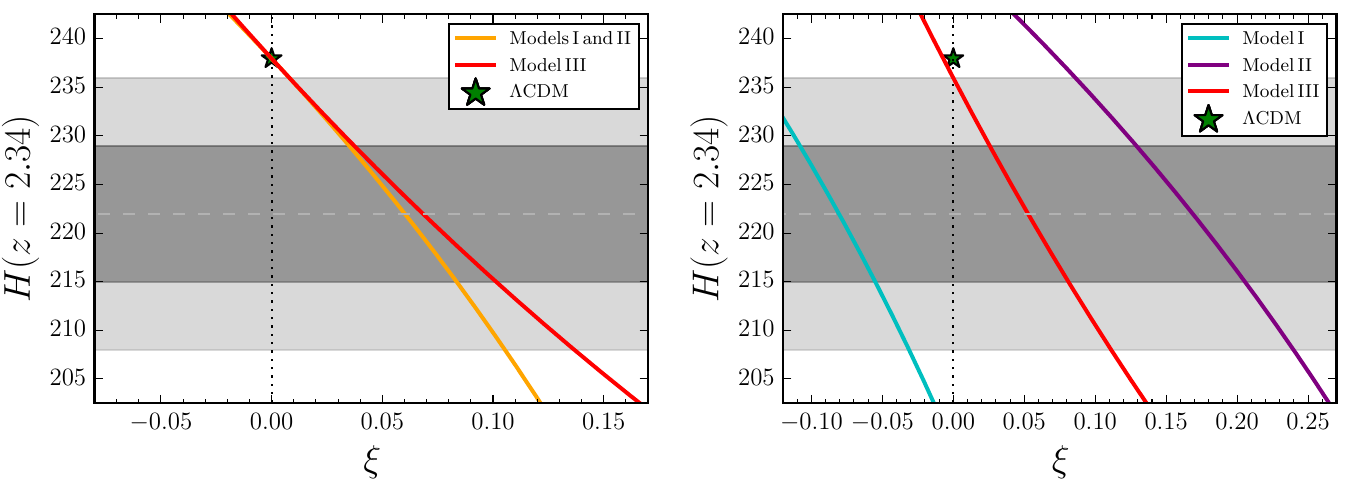}
\caption{We plot $H(z=2.34)$ as a function of the coupling $\xi$ (corresponding to $\xi_2$ for models I and II and to $\xi_1$ for model III).
The interacting models correspond to the colored lines since they depend on the free parameter $\xi$, the coupling constant.
The left panel represents the Hubble parameter calculated using the cosmological parameters from Table\ \ref{table:BOSS_parameters}
and with $\omega_\mathrm{DE}=-1$.
The right panel represents $H(2.34)$ using the parameters found in Ref.\ \cite{Costa:2013sva}
(including $\omega_\mathrm{DE}\neq -1$; see Table X for model I,
Table XI for model II, and Table XII for model III) obtained from Planck+BAO+SnIa+$H_0$.
The dashed gray line is the
BOSS measured value of $H(2.34)=222 \pm 7 ~\mathrm{km~s^{-1}~Mpc^{-1}}$, and the shaded
areas represent $1\sigma$ and $2\sigma$ deviations from this average.
For the sake of comparison, the green star represents $H(2.34)=238~\mathrm{km~s^{-1}~Mpc^{-1}}$,
the value expected for $\Lambda$CDM given the cosmological parameters in Table\ \ref{table:BOSS_parameters}.}
\label{fig:theoretical}
\end{figure*}

In order to compute the value of the Hubble parameter
from Eqs.~\eqref{H2},~\eqref{int_DE}, and~\eqref{int_DM},
one needs several cosmological parameters such as $H_0$,
$\Omega_\mathrm{DE}^{0}$, $\Omega_\mathrm{DM}^{0}$, and
$\Omega_\mathrm{b}^{0}$.
The standard $\Lambda$CDM parameters found from the Planck analysis
were used by the BOSS Collaboration (listed in Table\ \ref{table:BOSS_parameters}).
We first use these parameters and the dark energy EoS set to\footnote{The interacting models
are not well defined at the perturbative level if $\omega_\mathrm{DE}=-1$,
so we view $\omega_\mathrm{DE}=-1$ as a limit in this case.} $\omega_\mathrm{DE}=-1$ to construct $H(z)$, and we show
the resulting Hubble parameter at $z=2.34$ with respect to the coupling constant $\xi$ in
the left panel of Fig.~\ref{fig:theoretical}.
Alternatively, in the right panel of Fig.~\ref{fig:theoretical}, we use the adjusted cosmological parameters
found in Ref.\ \cite{Costa:2013sva} (including $\omega_\mathrm{DE}\neq -1$) from the analysis of the interacting models
using Planck, BAO, type Ia supernovae (SnIa), and $H_{0}$ data.
The goal of using different sets of cosmological parameters is to see if the parameters adjusted to the interacting models
yield a different prediction than the parameters adjusted to $\Lambda$CDM.

\begin{table}[htb]
\centering \caption{Cosmological parameters
used by the BOSS Collaboration
\cite{Delubac:2014aqe}.}
\begin{tabular}{ccc}
 \toprule
  Parameter & Best fit & $\sigma$ \\
   \hline
    $h$ & $0.706$ & $0.032$ \\
    $\Omega_\mathrm{DM}^{0} h^{2}$ & $0.143$ & $0.003$ \\
    $\Omega_\mathrm{DE}^{0}$ & $0.714$ & $0.020$ \\
    $\Omega_\mathrm{b}^{0} h^2$ & $0.02207$ & $0.00033$ \\
    \lasthline
\end{tabular}
\label{table:BOSS_parameters}
\end{table}

We recall that the BOSS Collaboration measured $H(2.34)=222 \pm 7 ~\mathrm{km~s^{-1}~Mpc^{-1}}$,
and this is indicated by the dashed gray line and by the $1\sigma$ and $2\sigma$ shaded areas
in Fig.~\ref{fig:theoretical}.
In comparison, standard $\Lambda$CDM cosmology predicts $H(2.34)\approx 238~\mathrm{km~s^{-1}~Mpc^{-1}}$
when using the cosmological parameters of Table\ \ref{table:BOSS_parameters}. This is represented by the
green star in Fig.~\ref{fig:theoretical}, which lies outside the $2\sigma$ measurement from BOSS.
In the left panel of Fig.~\ref{fig:theoretical}, all the curves that correspond to interacting dark energy pass through the green star at
$\xi=0$. This is because when the coupling constant vanishes there is no interaction left, and we recover $\Lambda$CDM
(since we set $\omega_\mathrm{DE}=-1$).
We also note that model I and model II correspond to the same curve,
because in the limit where $\omega_\mathrm{DE}=-1$, they correspond to the same model (recall Table\ \ref{table:models}).
In the right panel, we see that allowing for $\omega_\mathrm{DE}$ different than $-1$ can significantly alter
the prediction for $H(z=2.34)$.
Yet, all the curves can be in accordance with the Hubble parameter
inferred by BOSS given a nonzero coupling constant.
Comparing the left and right panels for model I, we notice that different cosmological
parameters require a different sign for the coupling constant $\xi$ in order to match the BOSS result.
This indicates that model I may not be fully robust at explaining the observed value of $H(z=2.34)$ from BOSS.
For models II and III, we see that
the theory can easily be within the $1\sigma$ shaded area
for a positive coupling constant in both panels.
We notice that in order for the $H(2.34)$ theoretical value to match the BOSS measurement, the values
of the coupling constant have to be larger in the right panel where the cosmological parameters
were adjusted to Planck+BAO+SnIa+$H_0$ data using the interacting models.

At this point, Fig.~\ref{fig:theoretical} provides us with indications that a positive coupling constant
allows one to explain in a very simple way a smaller value of the Hubble parameter in the past,
which is not possible with $\Lambda$CDM or dynamical dark energy and
without requiring a very exotic dark energy component.
The fact that we obtain a positive coupling constant for some models is interesting,
since it is precisely positive values that help alleviate the coincidence problem.
Thus, this model gives a natural explanation for the energy densities of the dark components at low redshifts
and also at high redshifts since they may explain the BOSS data.

This gives us evidence that the interacting dark energy model has the required features to be able to explain the different cosmological evolution
shown by the BOSS Collaboration at higher redshifts. However, this difference from $\Lambda$CDM dynamics is also encoded in the
angular distances, as inferred by the BAO measurement. We now compare the results for these parameters by
performing a global fit analysis of the interacting model with the currently available data.

\section{Analysis}

\subsection{Methodology}

Now that we see some evidence that the interacting dark energy models can explain the deviation from $\Lambda$CDM observed by BOSS,
we perform a Bayesian statistical analysis of those models with the Planck and BOSS Ly-$\alpha$
quasar data. We wish to compare the interacting dark energy models presented here against $\Lambda$CDM and test their predictions
with the addition of the new BOSS data. In order to achieve this, we perform a global fit by running the CosmoMC
package \cite{Lewis:2002ah}, a publicly available code that performs an MCMC parameter sampling.
To include the interaction between dark energy and dark matter, we modify the Boltzmann code CAMB \cite{Lewis:1999bs}
by adding the coupling constants $\xi_2$ for
model II and $\xi_1$ for model III and by adding the constant dark energy EoS to the
baseline $\Lambda$CDM parameters used by Planck \cite{Ade:2013zuv}.
From now on, we will omit model I from the analysis since this model showed us it was not very good to explain the new BOSS data.
Also, this model does not help alleviate the coincidence problem.
Model I will be explored in more detail in a follow-up paper.

The goal of this work is to compare the results of our global fit of
the cosmic distances and expansion rates for the interacting models with
the results obtained by the BOSS Collaboration. We also want to derive parameter constraints
using cosmic microwave background (CMB) and BAO data, testing the sensitivity of the parameters and in the total
goodness of fit when we include the new BAO data from higher redshifts.
The novelty of this work is in the BAO data that we use. The BOSS Collaboration was
the first team to measure the BAO from the autocorrelation of the quasar Ly-$\alpha$ forest
for higher redshifts. We use the autocorrelation  measurements from the DR11 catalog
from the BOSS experiment of SDSS which contains $158,401$ quasars in the redshift range
$2.1 \leq z \leq 3.5$\ \cite{Delubac:2014aqe}. From the same volume, cross-correlation of
quasars with the Ly-$\alpha$ absorption forest\ \cite{Font-Ribera:2013wce} was obtained for
the same redshift range. We are able to use both sets of data, since those can be considered as
independent, given that the fluctuations in the measurements are dominated by different
sources of systematics and not by cosmic variance. This analysis can be made by using the
baofit software provided by the BOSS Collaboration and the $\chi^2$ surfaces provided
for each one of those measurements\footnote{Available at \texttt{http://github.com/deepzot/baofit/}.}. 

For our global fit of the interacting dark energy models, we used the Planck $2013$ \textit{TT} power spectrum
in both the low-$\ell$ ($2 \leq \ell < 50$) and high-$\ell$ ($50 \leq \ell\leq 2500$) regimes.
Together with the Planck data, we include the polarization measurements from the nine-year
WMAP\ \cite{Bennett:2012zja},
the low-$\ell$ ($\ell < 32$) \textit{TE}, \textit{EE}, and \textit{BB} likelihoods.
In our first analysis, to illustrate the tension in the distance measurements between the BOSS measurement
and our global fit using Planck data, we combine the autocorrelation and cross-correlation $\chi^2$
surfaces provided by the BOSS Collaboration.

We also perform a joint analysis, where we include in the CosmoMC analysis the likelihood of the BOSS quasar Ly-$\alpha$ forest at
$z=2.34$. We can combine this new BAO data set with the CMB data sets since they are completely independent.
This was made in a very conservative way by inserting the two sets of Gaussian likelihoods constructed with the
best fit values of $(D_\mathrm{A} (z=2.34)/r_\mathrm{d},D_H (z=2.34)/r_\mathrm{d})$ for the autocorrelation and
cross-correlation given in Refs.\ \cite{Delubac:2014aqe,Font-Ribera:2013wce}. This appears to be a good choice, given that
the study of BAO from Ly-$\alpha$ is a novel field\footnote{Although this is a novel field, Ref.\ \cite{Delubac:2014aqe}
claims that the results are robust according to a consistency check using mock catalogs.}.

We used flat priors within the Planck $2013$ ranges for all the ``vanilla'' $\Lambda$CDM parameters\ \cite{Ade:2013zuv}.
The coupling constants\footnote{The coupling constants are expected to be small and positive, for models II and III,
from the previous analysis of Ref.\ \cite{Costa:2013sva}.
This was also indicated by the analysis in Fig.\ \ref{fig:theoretical}. These results motivated our choice of priors
for the interacting dark energy parameters.}
and dark energy EoS also received flat priors with $\xi_2 \in [0, \, 0.4[$ for model II,
$\xi_1 \in [0, \, 0.01]$ for model III, and $\omega\in[-2.5,\,-1.001]$ for both models.
We recall that we cannot allow for $\omega=-1$ since this represents a singularity in the perturbation equations.
The priors are summarized in Table\ \ref{table:priors}.

\begin{table}[htb]
\centering \caption{Priors for the parameters of the interacting dark energy models.
We recall that the definition of the different models is summarized in Table\ \ref{table:models}.}
\begin{tabular}{ccc}
 \toprule
  Model & Prior on $\omega$ & Prior on $\xi$ \\
   \hline
    II & [-2.5\,,\ -1.001] & [0\,,\ 0.4[ \\
    III & [-2.5\,,\ -1.001] & [0\,,\ 0.01] \\
    \lasthline
\end{tabular}
\label{table:priors}
\end{table}

\subsection{Results}

\begin{figure*}
 \centering
\includegraphics[width=0.45\textwidth]{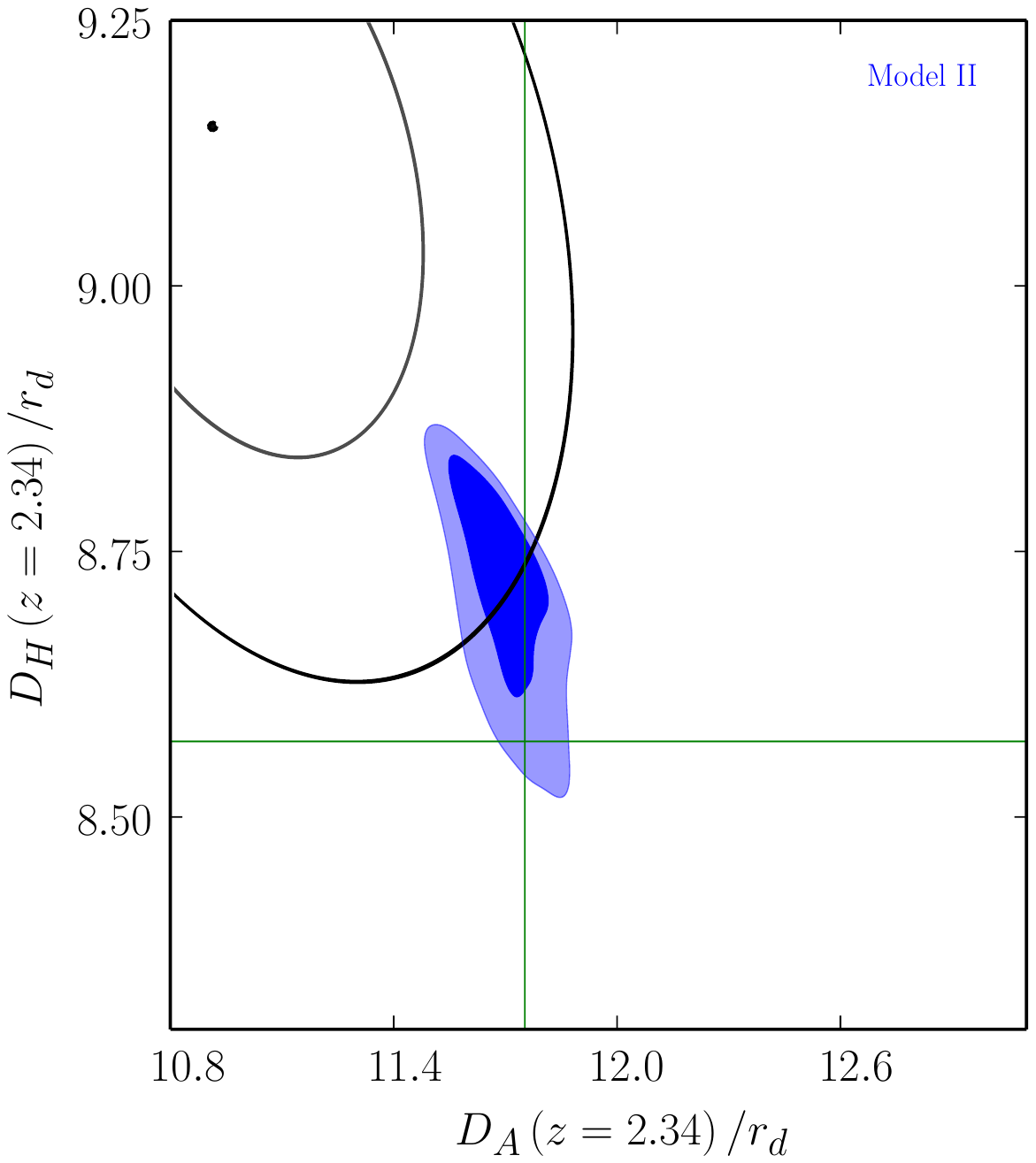}
\includegraphics[width=0.45\textwidth]{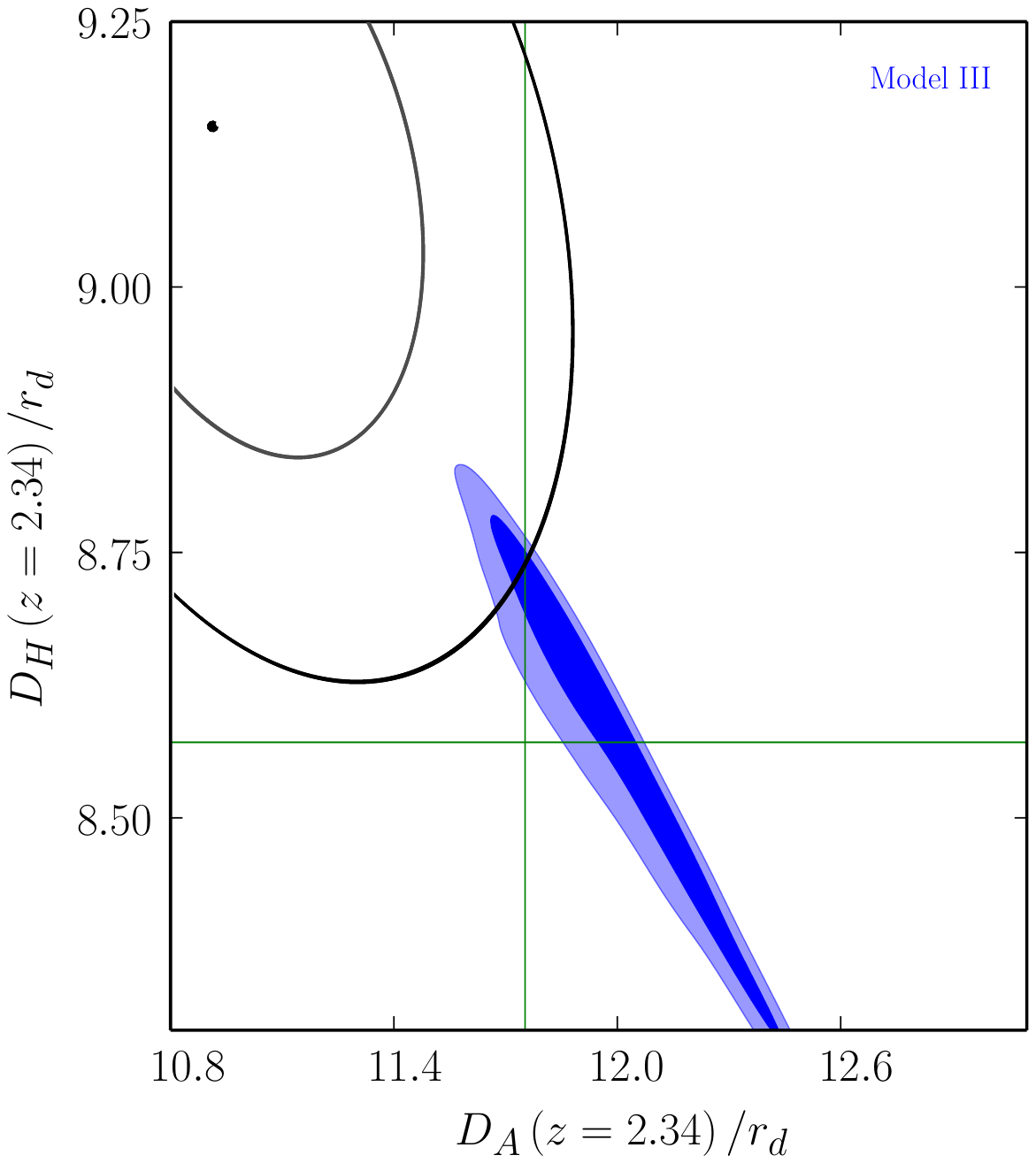}
\vspace{-1.3cm}
 \caption{Plot of the $68.3\%$ and $95.5\%$ likelihood contours
 in $D_\mathrm{A} (z=2.34)/r_\mathrm{d} \times D_H (z=2.34)/r_\mathrm{d}$ comparing the BOSS combined
 (autocorrelation and cross-correlation) contour in black
 with the results for the interacting models from the runs using Planck data in blue.
 Interacting model II is shown in the left panel and model III in the right panel.
 The green lines show the best fit values for $\Lambda$CDM.}
 \label{contours}
\end{figure*}
  
We wish to compare the constraints in $D_\mathrm{A} (z=2.34)/r_\mathrm{d}\times D_H (z=2.34)/r_\mathrm{d}$
found by the BOSS Collaboration with the global fits of the interacting dark energy models.
We present these results in Fig.~\ref{contours}. The black contour curves show the combined contours from the BOSS data
for the autocorrelation and cross-correlation\footnote{These contours are the same as the black
contour curves that one can find in Fig.~13 of Ref.\ \cite{Delubac:2014aqe}.}, given that those data are independent.

First, we perform the analysis using only CMB data for the $\Lambda$CDM
and interacting dark energy models. The constraints are shown by the blue contours in Fig.~\ref{contours} for models II and III.
We show for comparison the $\Lambda$CDM best fit values (green lines), where we obtain results compatible with Ref.\ \cite{Delubac:2014aqe},
which confirms that $\Lambda$CDM differs from the BOSS combined contours by at least $2\sigma$.
When we test the interacting models (blue contours),
this difference is reduced, and we can see that the contours overlap with the $2\sigma$ region of the BOSS combined data.
Model II, for which we find\footnote{Best fit values are presented inside brackets.}
$D_H/r_\mathrm{d}=8.72 (8.73)_{0.05}^{0.09}$ and $D_\mathrm{A}/r_\mathrm{d}=11.69(11.63) \pm 0.08$,
shows the biggest overlap with the BOSS results ($1.5\sigma$ and $1.7\sigma$ for $D_H/r_\mathrm{d}$ and $D_\mathrm{A}/r_\mathrm{d}$,
respectively).
The very elongated contours of model III imply that this conclusion is less strong in this case.

Although we show an apparent better concordance in comparison with the marginal overlap that $\Lambda$CDM presents
for $D_\mathrm{A} (z=2.34)/r_\mathrm{d}\times D_H (z=2.34)/r_\mathrm{d}$, this does not represent an improvement
in the fit, since the addition of extra parameters in the model can be the responsible for that. We can see the
same type of not-statistically-significant improvement for $\omega$CDM and other dynamical dark energy models
in Ref.\ \cite{Aubourg:2014yra}.
If you compare the constraints of our model II with the ones for $\omega$CDM at $z=2.34$
(see Fig.~7 of Ref.\ \cite{Aubourg:2014yra}), you can see that those contours almost overlap, showing a similar concordance with the new BOSS data.

Following that, we perform a joint analysis of the BOSS quasar Ly-$\alpha$ data together with the CMB data.
We wish to compare the improvement of the fit when including the new BOSS data. 
Our results indicate that $\Lambda$CDM is not sensitive to the inclusion of this data set (BOSS quasar Ly-$\alpha$ data), and therefore
it cannot accommodate the change in the Hubble parameter at high redshift. This shows a tension between those data sets.

The global fit of all the parameters of the interacting models reveals that the best fit values of the six vanilla
$\Lambda$CDM parameters are compatible with the ones obtained by Planck\ \cite{Ade:2013zuv},
except for model I, where the values for the density of matter show they are not in agreement with the Planck value.
We use $\Delta\chi^2_\mathrm{eff}$ to quantify the improvement in the maximum likelihood of the interacting dark energy models
using only Planck data in comparison to when we combine it with the likelihood from the BOSS team quasar data.
We found $\Delta \chi^2_\mathrm{eff}$ to be $-0.04$, $-2.88$, and $-1.85$, for models I, II and III, respectively.
Although these improvements are not statistically significant, they indicate that the interacting models,
and especially model II, are mildly favored by the data. Another test that also shows that the improvement
between the runs is not statistically significant is the reduced $\chi^2$, computed for all models.
This test takes into account that the interacting dark energy models have two extra degrees of freedom,
in comparison with the $\Lambda$CDM model. The difference in the reduced $\chi^2$ between
the interacting models and $\Lambda$CDM is not significant; e.g., model II presents the biggest
``improvement'' of the order of $10^{-5}$. However, one needs to be very careful when using an improvement
diagnostic like $\Delta \chi^2_\mathrm{eff}$ since the best fit values in CosmoMC may not be fully
trustworthy and since this result could come from statistics overfitting the noisy data \cite{Planck:2013jfk}.

\begin{figure}[htb]
\centering
\includegraphics[scale=1.1]{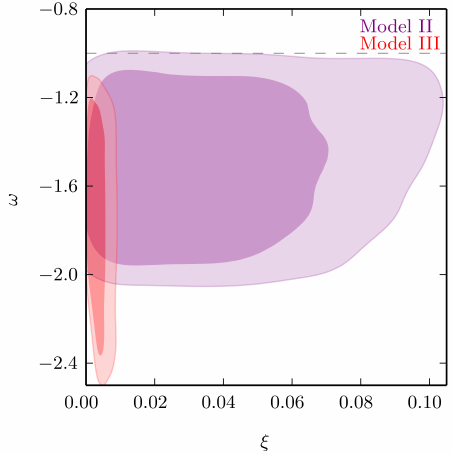}
\caption{Contour plot of the EoS for dark energy ($\omega$) vs the coupling constant between dark energy and dark matter ($\xi$).
In purple, we present the interacting model II, and in gray, we present the interacting model III fitted to the Planck data.
The cosmological constant $\Lambda$ of $\Lambda$CDM corresponds to $\omega = -1$, and it is depicted by
the dashed black horizontal line.}
\label{fig:w_xi}
\end{figure} 

In the MCMC analysis of the interacting models, we also obtained the adjusted values of the coupling constants.
As was shown in Ref.\ \cite{Costa:2013sva}, using only the Planck data is not sufficient to fully constrain the coupling constants.
We note that we obtain the same result here, even with the inclusion of the BOSS quasar data:
we find $\xi_2<0.045\, (0.048)$ for model II
and $\xi_1<0.0016\, (0.0015)$ for model III.
The upper bound on the coupling constant for model II is close to the ones predicted in Sec.~II-B (see Fig.~\ref{fig:theoretical}).
Indeed, the corresponding Hubble parameters that result from the MCMC analysis
are $H(2.34)= 232(231) \pm 2 \, \mathrm{km/s/Mpc}$ for model II
and $H(2.34)= 234(234)^2_3 \, \mathrm{km/s/Mpc}$ for model III, a little bit more than $1\sigma$
away from the BOSS result\footnote{We would like to stress that $H(z)$ is a model-dependent quantity, while $D_H/r_\mathrm{d}$ is not.
It is in this context that we compare our results with BOSS.
However, since we find that the fitted values for $r_\mathrm{d}$ are approximately equal
to what one expects in $\Lambda$CDM (given the use of the Planck data),
we can still compare the Hubble parameter values for the interacting models with the BOSS result.},
resulting in a reduced tension compared to $\Lambda$CDM.
This indicates that the interacting models are good candidates to explain the observed deviation from $\Lambda$CDM from high-$z$ BAO probes. 
The upper bound on the coupling constant for model III is much smaller than expected from Fig.~\ref{fig:theoretical}.
Still, it represents an improvement over $\Lambda$CDM in explaining the BOSS results as seen from Fig.~\ref{contours},
although to a smaller extent than model II.

The upper bounds found for the coupling constants are compatible with small positive values.
Although we cannot exclude the possibility that the coupling constants are zero with the data set used, we can see from the constraints obtained
for the EoS of dark energy that our models are not consistent with $\Lambda$CDM.
The EoS for dark energy obtained in the MCMC analysis are the following: considering only Planck data,
$\omega = -1.51(-1.55)^{+0.32}_{-0.30}$ for model II and $\omega=-1.75(-1.668)^{+0.46}_{-0.29}$
for model III. We can also see the constraints
in the $\omega \times \xi$ plot, presented in Fig.~\ref{fig:w_xi}. The dashed black horizontal line represents the value of the dark energy
EoS for $\Lambda$CDM, $\omega =-1$.
These contours show a small preference for $\omega<-1$ rather than $\omega=-1$ given the priors, $\omega = [-2.5\,,\, -1.001]$, with model
II showing a slightly tighter constraint than the prior range.
This result should be interpreted carefully since our prior is very close to $-1$ (but it is not including $-1$),
and there can be boundary effects that might not be taken into account.
Also, we have a large degeneracy between $\omega$ and $\xi$.

A more detailed analysis will be presented in a follow-up paper where we will combine this analysis with different cosmological probes,
aiming at fully constraining the coupling constant of the interacting models.

\section{Conclusions}

In this paper, we explored the consequences of interacting dark
energy in light of the recent results by the BOSS experiment.
The BOSS data indicate that the Hubble parameter at $z=2.34$ is smaller than
what one would expect from the standard $\Lambda$CDM model,
something that cannot be explained by simple dynamical dark energy models
such as quintessence. Our results suggest that interacting dark energy can
naturally explain the BOSS data without introducing exotic forms of dark energy., although further studies are necessary.

We tested three different phenomenological models
of interacting dark energy.  First, we computed the theoretical value of the Hubble parameter
at $z=2.34$ for different sets of cosmological parameters.
Models II and III showed they were in good agreement with
the observations for a small positive coupling constant. 
Furthermore, such a positive coupling constant can help
alleviate the coincidence problem.
Model I was omitted from the analysis since it did not
contribute to reducing the tension with the BOSS data,
and also, in general, it does not help relieve the coincidence problem.

We then performed a global fit of those models given the Planck 2013 and BOSS quasar Ly-$\alpha$ data.
This showed that models II and III present a bigger overlap with the BOSS Collaboration results than what $\Lambda$CDM achieves.
However, this improvement and also the improvement in the $\chi^2$ when we made the joint analysis with CMB and BOSS likelihoods
do not seem to justify the inclusion of extra parameters in the model as done by the interacting models. In this analysis,
we can also see from the EoS obtained that those models are marginally different than $\Lambda$CDM.
Yet, the results still suggest that the interacting dark energy models presented in this paper can be used to explain the deviations
from $\Lambda$CDM found in high-$z$ BAO, and they represent a simpler solution than invoking exotic dark energy models. 

In order to further constrain interacting dark energy models,
one could refine the analysis done in this work by using more data sets
and by combining the BOSS data with other observations. 
A more detailed analysis of the global fit of those models with the inclusion of BOSS data 
is the topic of a follow-up paper that is currently in preparation.
We also need improvements in the BAO data at high redshifts.
For models that allow the Hubble parameter to change with time such as interacting dark energy and other dynamical dark
energy models (e.g., see Ref.\ \cite{Aubourg:2014yra}), we can see that the inclusion of the BAO
data set changes considerably the results, indicating that this new data set is robust. However, with the use of
only high-redshift BAO data, we are still not able to statistically differentiate between models of dark energy.
New large scale structure surveys, like the JPAS telescope\ \cite{Benitez:2014ibt}, will be able to reproduce
and improve the BAO measurements at high redshifts since
this instrument is supposed to be optimized to measure quasars
at high redshifts compared to previous experiments\ \cite{Abramo:2011ey}.
Other large scale structure new windows of observation,
like the $21$ cm emission line from neutral hydrogen, will also contribute in the
future for constraining dark energy \cite{Battye:2012tg}.
Interacting dark energy models might also help alleviate the tension between
other large-scale structure data sets and Planck such as, for example,
cosmic shear probes from CFHTLenS\ \cite{Joudaki:2016mvz,MacCrann:2014wfa}.

\begin{acknowledgments}
We thank Gil Holder for useful discussions.
E.\,F. and E.\,A. acknowledge financial support from 
Conselho Nacional de Desenvolvimento Cient\'\i
fico e Tecnol\' ogico (CNPq) and from Funda\c
c\~ao de Amparo \`a Pesquisa do Estado de S\~ao Paulo (FAPESP).
J.\,Q. acknowledges the support, throughout the completion
of this work, of the Fonds de Recherche du Qu\'ebec \textemdash\ Nature
et Technologies (FRQNT), the Walter C.~Sumner Foundation,
and the Natural Sciences and Engineering Research Council
(NSERC) of Canada via the Vanier Canada Graduate Scholarships program.
A.\,C. thanks FAPESP and Coordination for the Improvement of Higher Education Personnel (CAPES) for the financial support
under Grant No.\ 2013/26496-2 (FAPESP).
B.\,W. is supported by the National
Basic Research Program of China (973 Program No.\ 2013CB834900) and by the National Natural Science
Foundation of China.
This work has made use of the computing facilities of the Laboratory of Astroinformatics (IAG/USP, NAT/Unicsul),
the purchase of which was made possible by the Brazilian agency FAPESP (2009/54006-4) and the INCT-A.
Computations were also made in part on the supercomputer
Guillimin from McGill University, managed by Calcul
Qu\'ebec and Compute Canada. The operation of this supercomputer
is funded by the Canada Foundation for Innovation
(CFI), NanoQu\'ebec, RMGA, and FRQNT.
\end{acknowledgments}


\end{document}